\documentclass[aps,pra,twocolumn,floatfix]{revtex4-1} 
\usepackage{graphicx}
\usepackage{amssymb,amsmath}
\usepackage{color}
\usepackage{bbold}

\newcommand{\ket}[1]{|#1\rangle}				
\newcommand{\bra}[1]{\langle #1|}				
\newcommand{\ew}[1]{\langle #1 \rangle}				
\newcommand{\ketbra}[1]{| #1 \rangle \langle #1 |}		

\renewcommand{\vec}[1]{\boldsymbol{#1}}

\def\cdott{\!\cdot\!}
\def\dls{\delta_\text{LS}}
\def\Dls{\Delta_\text{LS}}
\def\gs{\gamma_\text{S}}
\def\zs{\zeta_\text{S}}
\def\kR{\kappa_\text{R}}
\def\kT{\kappa_\text{T}}
\def\dc{\Delta_\text{C}}
\def\aout{a_{\text{out}}}
\def\bout{b_{\text{out}}}
\def\ain{a_{\text{in}}}

\begin{document}
\title{X-ray quantum optics with M\"ossbauer nuclei embedded in thin film cavities}

\author{Kilian P. Heeg}

\author{J\"org Evers}

\affiliation{Max-Planck-Institut f\"ur Kernphysik, Saupfercheckweg 1, 69117 Heidelberg, Deutschland}

\date{\today}
\begin{abstract}
A promising platform for the emerging field of x-ray quantum optics are M\"ossbauer nuclei embedded in thin film cavities probed by near-resonant x-ray light, as used in a number of recent experiments. Here, we develop a quantum optical framework for the description of experimentally relevant settings involving nuclei embedded in x-ray waveguides. We apply our formalism to two settings of current experimental interest based on the archetype M\"ossbauer isotope $^{57}$Fe. For present experimental conditions, we derive compact analytical expressions and show that the alignment of medium magnetization as well as incident and detection polarization enable the engineering advanced quantum optical level schemes. The model encompasses non-linear and quantum effects which could become accessible in future experiments.
\end{abstract}


\maketitle

\section{Introduction}

X-ray quantum optics is a promising emerging field at the boundary of visible quantum optics and x-ray science~\cite{Adams_review}. A particularly interesting platform for the exploration of x-ray quantum optics are M\"ossbauer nuclei, which offer a number of unique features. Among them are their narrow resonances, which on the one hand enable the manipulation and observation of nuclei in the time domain, and on the other hand offer interesting perspectives for precision spectroscopy. Another feature is the abundance of cooperative effects, as nuclei are commonly probed in solid state targets with large numbers of nuclei. A recent experiment could also demonstrate that nuclei can be operated essentially decoherence-free~\cite{heeg_sgc}.

Recently, a number of prominent quantum optical effects could be observed with nuclei, such as the cooperative Lamb shift~\cite{Roehlsberger_Lamb_Shift}, electromagnetically induced transparency~\cite{Roehlsberger_EIT}, and spontaneously generated coherences~\cite{heeg_sgc}. Also the possibility to dynamically control the light-matter interaction has already been demonstrated, e.g., by rapid switching of applied magnetic fields~\cite{Shvydko_Switching}, or by dynamic modifications of the sample geometry~\cite{Kocharovskaya_Switching}. Quantum mechanical aspects have been touched, e.g., in first experiments on x-ray photon downconversion~\cite{adams_nlo,Eisenberger_XPDC,Yoda_XPDC}, and also alternative methods to generate x-ray entanglement have been proposed~\cite{palffy_entanglement}.

These examples illustrate that M\"ossbauer and thus x-ray science can profit from well-established ideas developed in the visible frequency range. But it is important to realize that quantum optics as a whole can profit from the progress in x-ray science equally well. None of the above recent examples relied on a simple transfer of setups from the optical to the x-ray frequency range. Instead, new ideas and techniques had to be developed, which potentially could be ported back to the optical frequency range. 

While nuclei can in principle directly be driven with x-ray light sources~\cite{Burvenich_xrqo,Burvenich_Stark}, a significant part of recent progress in x-ray quantum optics with M\"ossbauer nuclei has been enabled using nuclei embedded in thin-film cavities probed in grazing incidence by hard x-rays. The cavity is formed by a stack of thin layers made from different materials, such that differences in the refractive index lead to the formation of the waveguide structure. Similar waveguides have been studied in the context of light propagation and focusing of x-rays  \cite{Salditt_High_T_Xray_Wavguides,Pfeiffer_Xray_Waveguides}. For the theoretical description of the optical properties of such cavities including M\"ossbauer nuclei, a matrix formalism has been developed, which self-consistently treats the scattering between the different layers~\cite{Roehlsberger_grazing_incidence_theory,Roehlsberger_Scattering}. A numerical variant of this formalism is implemented in the software package CONUSS \cite{Sturhahn_CONUSS} and is considered as benchmark for other theories, as it has proven to agree very well with experimental data. 
In the particular case of thin resonant layers of nuclei embedded in the waveguide, analytic expressions for the cavity properties can be obtained. This approach formed the basis for the interpretation recently observed quantum optical effects~\cite{Roehlsberger_Lamb_Shift,Roehlsberger_EIT,heeg_sgc}. 
This invites further study of more complex nuclear waveguide systems, which prompts for more powerful theoretical descriptions.

Motivated by this, here, we ab initio develop a quantum optical framework for the modeling of large ensembles of nuclei embedded in thin film cavities and probed in grazing incidence by hard x-rays. We start with the derivation of a master equation for the full ensemble of nuclei coupled to the quantized cavity modes. We include all magnetic sublevels, such that arbitrary alignments of the magnetization as well as the input- and output polarization can be analyzed. The model includes non-linear and quantum effects, which could become accessible in future experiments. Motivated by the present experimental state-of-the-art, we then specialize to the case of lossy cavities and linear response. This allows us to derive analytic solutions by adiabatically eliminating the cavity modes and by characterizing the large ensemble of nuclei using few many-body quantum states. As a main result, we find that the considered setup enables us to engineer a wide range of few-level quantum optical systems in the x-ray regime, with level structure tunable via the applied magnetization and the light polarizations. The corresponding master equation allows to fully identify and interpret all physical mechanisms contributing to the obtained results. 
Finally, we focus on the most relevant case of ${}^{57}$Fe, and illustrate our framework by analyzing two settings of current experimental interest. The first one is the simplest setting of a single unmagnetized layer of nuclei placed in the center of an x-ray cavity. Consistent with recent experimental results, our analysis predicts cooperative Lamb shifts and superradiance. Second, we consider a single layer including magnetic hyperfine splitting, such that the spectrum in general consists of six transition lines. 
We find that our approach is analytically equivalent to existing approaches in the respective limits. But it goes beyond the existing approaches by opening perspectives for the engineering of advanced quantum optical schemes in the hard x-ray regime. It enables the generalization to cases in which the quantum nature of the x-ray light is of relevance as, e.g., in quantum information theory. Moreover, it can cover situations in which the light source delivers many resonant photons per shot, such that non-linear effects become crucial, and offers full interpretation in terms of the involved physical processes.

\section{Setup}
\begin{figure}[bt]
 \centering
 \includegraphics[scale=0.8]{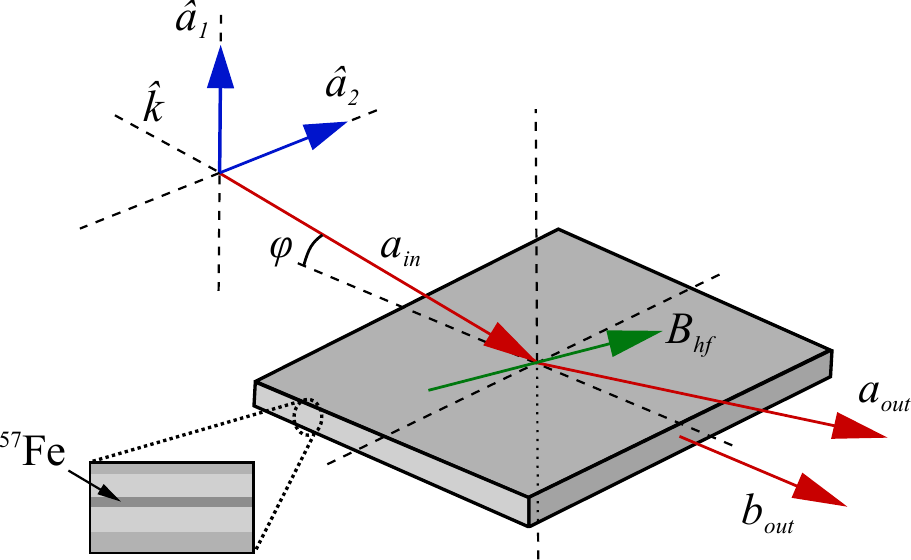}
 \caption{(Color online) Schematic of the considered setup. The cavity contains a layer of resonant nuclei as indicated in the inset. It is probed by hard x-rays (red lines, $\ain$) with propagation direction $\vec{\hat k}$. The angle of incidence $\varphi$ is of the order of a few mrad. The incident polarization in the ($\vec{\hat a}_1,\vec{\hat a}_2$) plane (blue) together with the alignment of the magnetization $\vec{B}_\text{hf}$ of the nuclei (green) sensitively determine the properties of the scattered light. Both, light reflected from the cavity ($\aout$) at output angle $\phi$ and light exiting the cavity on the front side ($\bout$) are considered. }
 \label{fig:cavity}
\end{figure}

\begin{figure}
 \centering
 \includegraphics[width=0.9\columnwidth]{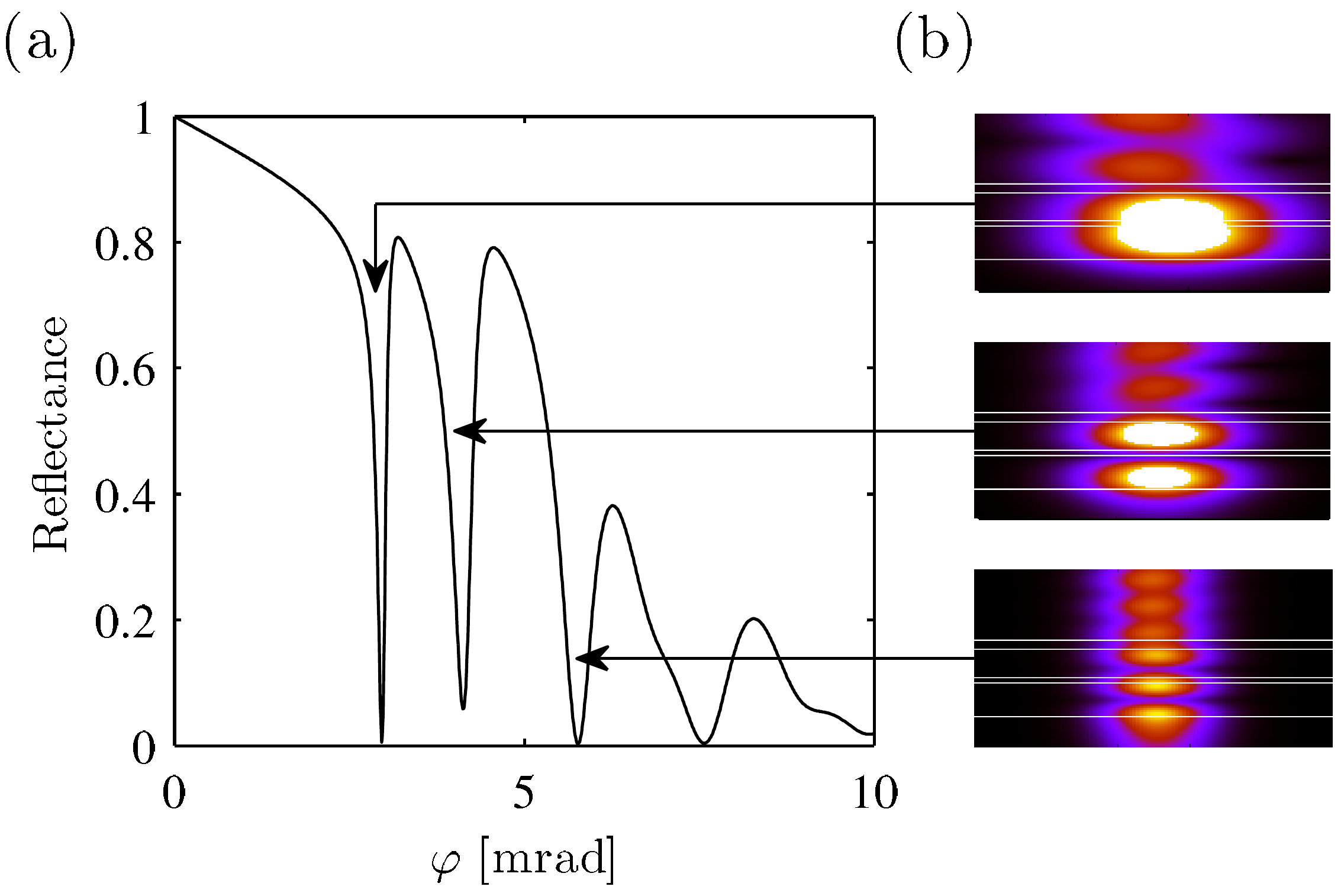}
 \caption{(Color online) (a) Electronic reflectivity curve of a thin film cavity, showing the reflectance as a function of the x-ray incidence angle $\varphi$. The dips correspond to resonant excitations of guided modes of the waveguide. The cavity consists of a 2.6 nm Pt top layer, followed by 7.9 nm C, 1.5 nm $^{57}$Fe, 9.3 nm C, and a thick bottom layer of Pt. (b) Field intensity distributions inside the waveguide (without Fe layer for simplicity) for probing fields resonant to the first three guided modes, respectively. The three panels only differ in the incidence angle of the field, and show a range of 1~mm in horizontal and of 50~nm in vertical direction. The white horizontal lines indicate the layer boundaries, and above the layer system, the standing wave formed by incident and reflected field is visible.}
 \label{fig:rocking}
\end{figure}

\subsection{Cavity}
The system we investigate in this work is a thin film cavity probed by hard x-rays as shown in Fig.~\ref{fig:cavity}~\cite{Roehlsberger_Scattering}. The thin film typically consists of layers of different materials with thicknesses of the order of a few nanometers. On one hand, the probing incident light indicated by the field $\ain$ in Fig.~\ref{fig:cavity} can be reflected from the layer structure, with outgoing light indicated by photon operator $\aout$. On the other hand, the layer structure can be chosen in such a way that a cavity or waveguide is formed for the probing light. This is achieved by combining materials with low electron density (e.g., carbon) in the center of the structure, and materials with high electron density (e.g., platinum or palladium) at outer layers which act as mirrors. The electron density translates into the index of refraction experienced by the probing x-ray light. The spatial modulation of the index of refraction leads to reflection of the light at the boundaries, resulting in a waveguide. In this case, the probing light in addition can evanescently couple into waveguide modes, and eventually exit the layer structure to the side, as indicated by photon operator $\bout$ in Fig.~\ref{fig:cavity}. Note that in contrast to the optical regime, in the x-ray regime the real parts of the index of refraction are typically below one \cite{Roehlsberger_Scattering}, such that a low electron density in the center leads to guiding of the light, together with total external reflection.

Because of the small index of refraction variations at x-ray energies, the cavity is typically probed at grazing incidence, with small incident angle $\varphi$ with respect to the cavity surface, as shown in Fig.~\ref{fig:cavity}. The reflectance and the coupling into waveguide modes sensitively depends on this angle, as illustrated in Fig.~\ref{fig:rocking}(a). The dips in this electronic reflectivity curve arise if the angle $\varphi$ leads to resonant coupling of the probing light into a particular waveguide mode. The field intensity distribution in the waveguide is illustrated in Fig.~\ref{fig:rocking}(b) for the first three guided modes. 

In an experiment, the light and its properties reflected from the cavity ($\aout$) or transmitted through the cavity ($\bout$) can be recorded. In general, both the energy and the time spectrum of the outgoing light will strongly be modified by the interaction with the cavity.

\subsection{\label{sec:nuclei}Nuclei}

So far, we have only discussed the properties of the waveguide in terms of electronic scattering of the x-rays from the materials of the layer system. For the purpose of x-ray quantum optics, in addition layers of nuclei can be embedded into the waveguide~\cite{Roehlsberger_Lamb_Shift,Roehlsberger_EIT,heeg_sgc}. Such layers contain a large ensemble of nuclei, which can coherently interact with the probing x-ray light entering the waveguide. By carefully choosing the position of the nuclei inside the layer structure, as well as the resonantly driven mode of the waveguide [see Fig.~\ref{fig:rocking}(b)], the interaction between nuclei and the light inside the cavity can be controlled. In particular, different layers of nuclei can interact in a different way with the same cavity mode~\cite{Roehlsberger_EIT}. 

\begin{figure}
 \centering
 \includegraphics[scale=0.6]{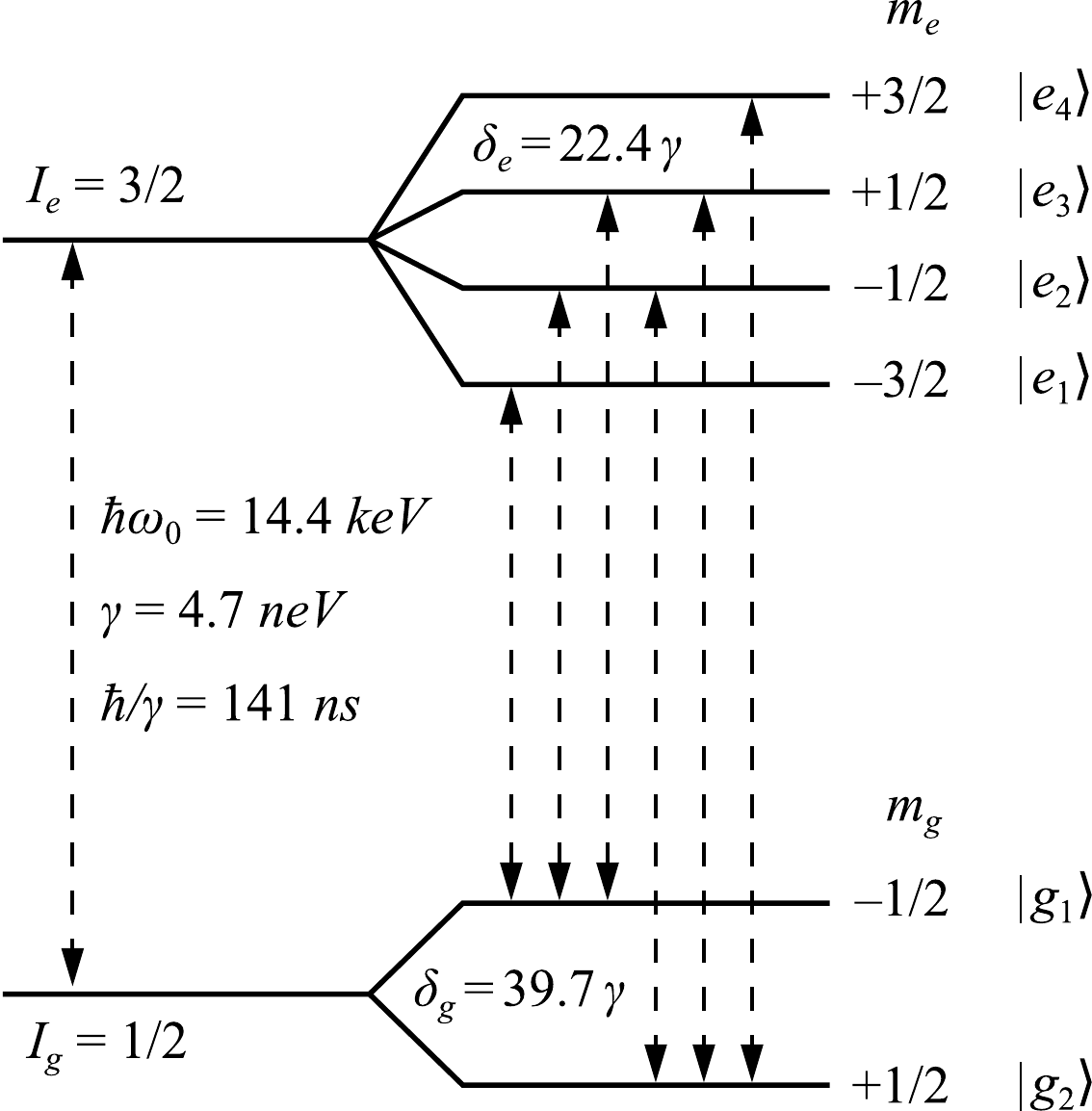}
 \caption{The M\"ossbauer transition in $^{57}$Fe. In the presence of a magnetic hyperfine field the two levels split up and six M1 transitions can be driven.}
 \label{fig:fe_scheme}
\end{figure}

In this work, we focus on the most frequently used archetype M\"ossbauer isotope $^{57}$Fe illustrated in Fig.~\ref{fig:fe_scheme}. This isotope features a transition from the ground state to the first excited state at $\omega_0 = 14.4$~keV with single-nucleus linewidth $\gamma=4.7$~neV ($\hbar = 1$ used here and in the following). In the absence of magnetic fields, it acts as a two-level system. In the presence of magnetic fields, the ground and excited states with $I_g = 1/2$ and $I_e = 3/2$ split into multiplets shown in Fig.~\ref{fig:fe_scheme}. In general, six different transitions between ground and excited states are possible. Note that the considered transition is a magnetic dipole (M1) transition, such that the polarization vectors to be defined later have to be identified with the magnetic polarizations of the incoming and outgoing radiation, respectively. We furthermore specialize to the case of a single layer of nuclei, which we place at a maximum of the field intensity distribution of the cavity, in order to maximize the nucleus-field interaction.

It is the inclusion of resonant nuclei which qualifies the considered system for applications in x-ray quantum optics. Close to nuclear resonances, the properties of the combined system of waveguide and nuclei lead to a strong polarization and energy dependence of the scattered light.

\section{Theoretical model}
\subsection{Cavity}

As pointed out in the previous part, the incident x-ray beam can resonantly couple to waveguide modes inside the cavity at particular values for the angle of incidence $\varphi$. It is instructive to characterize the modes in terms of the resonant cavity wave vector $\vec{k}_C$. First we note that the external x-ray field with frequency $\omega$ and wave vector $\vec k$ can be divided into components $k_z = |\vec k| \sin (\varphi)$ perpendicular and $k_x = |\vec k| \cos(\varphi)$ parallel to the surface. In order to satisfy the continuity relations of Maxwell's equations at boundaries, the parallel components $k_{Cx}$ inside and $k_x$ outside the cavity must be identical. In contrast, the perpendicular component $k_{Cz}$ of the mode is determined by parameters of the cavity such as the thickness of the layers and the refractive indices and the order of the guided mode \cite{Kane_GuidedModes,DeWames_ThinFilms}. This means the cavity possesses resonances only in the direction perpendicular to the surface, while the parallel components of total resonance wave vector can be chosen freely.
In this work we will restrict the discussion to only one guided mode and assume, without loss of generality, that its resonance condition for $k_{Cz}$ is fulfilled if an x-ray beam with the resonance frequency of the $^{57}$Fe transition $\omega_0 = c\cdot k_0$ impinges on the layer surface under an angle of incidence $\varphi_0$. In this case $k_{Cz} = k_z^0 = k_0 \sin (\varphi_0)$ and $k_{Cx} = k_x^0$, and we find that the cavity mode is resonantly driven. If the angle of incidence is varied from $\varphi_0$ to a general angle $\varphi$, the perpendicular mode component $k_{Cz}$ is still fixed by the same resonance condition $k_{Cz} = k_0 \sin (\varphi_0)$ of the waveguide mode, while $k_{Cx} = |\vec{k}| \cos(\varphi)$ can freely vary with $\varphi$. In other words, the mode of interest is not only characterized by cavity parameters, but also by the incident beam. For the total wave vector of the resonant cavity mode, this yields
\begin{equation}
 |\vec{k}_C| = \sqrt{|\vec{k}|^2 \cos{(\varphi)}^2 + k_0^2 \sin{(\varphi_0)}^2} \;.
 \label{eqn:wavevector_cavity}
\end{equation}

We now continue with the derivation of the Hamiltonian for this mode in the cavity and its driving due to the external field. In a first step, we do not yet take into account any polarization dependence. In the Schr\"odinger picture the Hamiltonian reads \cite{Scully_Zubairy_Quantum_Optics} 
\begin{equation}
 H_{M}^{(S)} = \omega_C a^\dagger a + i \sqrt{2 \kR} \left( a_\text{in} e^{-i \omega t} a^\dagger - a_\text{in}^* e^{i \omega t} a \right) \;.
\end{equation}
Here $a$ [$a^\dagger$] is the photon annihilation [creation] operator for the field in the cavity, $\ain$ characterizes the driving of the cavity mode by the external classical x-ray field with frequency $\omega$ and $\omega_C = c\cdot|\vec{k}_C|$ is the mode resonance frequency. In a next step we transform the system into an interaction picture to eliminate the explicit time dependence in the Hamiltonian. We apply the unitary transformation
\begin{equation}
 \ket{\Psi^{(I)}} = U^\dagger\ket{\Psi^{(S)}}
\end{equation}
given by
\begin{subequations}
\begin{align}
 U &= \exp({-i H_T t}) \,,\\
 H_T &= \omega \, a^\dagger a\,,
 \label{eqn:how_to_trafo}
\end{align}
\end{subequations}
and obtain the interaction picture Hamiltonian
\begin{align}
 H_M^{(I)} &= U^\dagger H_{M}^{(S)} U - H_T \nonumber \\
 &= \dc a^\dagger a + i \sqrt{2 \kR} \left( a_\text{in} a^\dagger - a_\text{in}^* a \right) \;. \label{eqn:Hamiltonian_Modes_simple}
\end{align}
Here we introduced the cavity detuning $\dc = \omega_C- \omega$. For a small angular deviation $\Delta \varphi = \varphi - \varphi_0$ from the resonant incident angle $\varphi_0$ and $\omega \approx \omega_0$ we find from Eq.~(\ref{eqn:wavevector_cavity})
\begin{align}
 \dc &= \sqrt{\omega^2\cos{(\varphi_0 + \Delta\varphi)}^2+\omega_0^2\sin{(\varphi_0)}^2} -\omega \nonumber \\
 &\approx - \omega \varphi_0 \Delta\varphi \;, \label{eqn:cavity_detuning}
\end{align}
such that the detuning is proportional to the incidence angle and the frequency of the incident light.

Now we generalize this Hamiltonian to the case including polarization. We denote the beam propagation direction as $\vec{\hat k}$, where the ``hat'' indicates a normalized unit vector. Since $\varphi_0 \ll 1$, the direction of the incident, reflected and transmitted beam can be considered as equal, parallel to $\vec{\hat k}$. As a consequence, their respective polarizations $\vec{\hat a}_\text{in}$, $\vec{\hat a}_\text{out}$ and $\vec{\hat b}_\text{out}$ are located in the plane defined by the layer surface normal $\vec{\hat a}_1$ and $\vec{\hat a}_2 = \vec{\hat a}_1 \times \vec{\hat k}$. Including both these polarizations as different modes $a_1$ and $a_2$ in our calculation, the Hamiltonian in the interaction picture becomes
\begin{align}
 H_M &= \dc {a_1}^\dagger {a_1} + \dc {a_2}^\dagger {a_2} \nonumber \\ 
     &+ i \sqrt{2 \kR} \left[ (\vec{\hat a}_1^* \cdott \vec{\hat a}_\text{in}) \, a_\text{in} a_1^\dagger - (\vec{\hat a}_\text{in}^* \cdott \vec{\hat a}_1) \, a_\text{in}^* a_1 \right] \nonumber \\
     &+ i \sqrt{2 \kR} \left[ (\vec{\hat a}_2^* \cdott \vec{\hat a}_\text{in}) \, a_\text{in} a_2^\dagger - (\vec{\hat a}_\text{in}^* \cdott \vec{\hat a}_2) \, a_\text{in}^* a_2 \right] \; .
     \label{eqn:Hamiltonian_Modes}
\end{align}
Here, $(\vec{\hat a}_i^* \cdott \vec{\hat a}_j)$ are scalar products between two different polarization unit vectors.

Next to the coherent dynamics described by Eq.~(\ref{eqn:Hamiltonian_Modes}), also incoherent processes need to be considered. This is particular important as in typical experiments, the cavity has a relatively low $Q$ factor \cite{Roehlsberger_Lamb_Shift}. It is important to note that incoherent processes such as spontaneous emission evolve a pure quantum mechanical state into an incoherent mixture of states, which cannot be described using a wave function. Therefore, we include incoherent processes using the master equation approach~\cite{Scully_Zubairy_Quantum_Optics,Kiffner_Vacuum_Processes} for the system's density matrix $\rho$. In this framework, the governing equation replacing the usual Schr\"odinger equation is 
\begin{equation}
 \frac{d}{dt}\rho = -i [H, \rho] + \mathcal{L}[\rho] \;,
\end{equation}
where the commutator part $[\cdot,\cdot]$ characterizes the coherent evolution by the Hamiltonian $H$, and the Lindblad operator $\mathcal{L}[\rho]$ models incoherent processes. For arbitrary operators $\mathcal{O}^+$ and $\mathcal{O}^-$, a contribution to the latter can be defined as
\begin{equation}
  \mathcal{L}[\rho, \mathcal{O}^+, \mathcal{O}^-] = \big( \mathcal{O}^+ \mathcal{O}^- \rho + \rho \mathcal{O}^+ \mathcal{O}^- -2\mathcal{O}^- \rho \mathcal{O}^+ \big) \label{eqn:lindblad_helper} \;.
\end{equation}
With this definition, the photon loss out of modes $a_1$ and $a_2$ can be written as \cite{Dayan_Photon_Turnstile}
\begin{align}
 \mathcal{L}_M[\rho] = &-\kappa\,  \mathcal{L}[\rho, a_1^\dagger, a_1] -\kappa\,  \mathcal{L}[\rho, a_2^\dagger, a_2] \label{eqn:cavity_decay} \;.
\end{align}
Note that cavity loss in the present framework not only arises due to incoherent scattering or absorption in the layer structure, but also by outcoupling of the cavity field into the modes characterizing reflectance and transmittance. The total rate $\kappa$ contains all of these loss processes. In the absence of nuclei, $ \mathcal{L}[\rho] = \mathcal{L}_M[\rho]$. With nuclei embedded in the cavity, further incoherent processes associated with the nuclei arise, which we discuss in Sec.~\ref{sec:inc_nuc}.

\subsection{\label{sec:io}Input-Output relations}
In an experiment not the internal modes in the cavity, but the reflected ($\aout$) or transmitted beams ($\bout$) are observed. These output field operators can be calculated using the input-output formalism~\cite{Gardiner_Quantum_Noise}. Assuming polarization-sensitive detection with detector polarization $\vec{\hat a}_\text{out}$ they read
\begin{align}
 a_\text{out} &= -a_\text{in} \, \left(\vec{\hat a}_\text{out}^* \cdott \vec{\hat a}_\text{in} \right) \nonumber \\
 &+ \sqrt{2 \kR} \left[ (\vec{\hat a}_\text{out}^* \cdott \vec{\hat a}_1) \, a_1 + (\vec{\hat a}_\text{out}^* \cdott \vec{\hat a}_2) \, a_2 \right] \,,\\
 b_\text{out} &= \sqrt{2 \kT} \left[ (\vec{\hat a}_\text{out}^* \cdott \vec{\hat a}_1) \, a_1 + (\vec{\hat a}_\text{out}^* \cdott \vec{\hat a}_2) \, a_2 \right] \; .
 \label{eqn:aout_bout}
\end{align}
Note that the transmission $\bout$ only receives contributions originating from the modes $a_1$ and $a_2$ inside the cavity, while $\aout$ also contains the part of the incident light $\ain$ directly reflected from the cavity. The coupling constant $\kR$ in Eq.~(\ref{eqn:aout_bout}) is equal to the corresponding one in Eq.~(\ref{eqn:Hamiltonian_Modes}), as both describe the coupling between the same internal and external modes. This parameter can be controlled by changing, e.g.~the thickness of the topmost layer. 
Further, we note that $\kappa \ge \kR + \kT$, because the cavity is not only damped by coupling into the outgoing modes, but also by internal loss, as discussed below Eq.~(\ref{eqn:cavity_decay}). This condition is crucial for fulfilling the energy conservation.

\subsection{Observables}
To guide the further analysis, it is useful to consider possible observables accessible in a typical experiment. These are primarily the reflectance (scattering into $\aout$ in Fig.~\ref{fig:cavity}) and the transmittance (scattering into $\bout$ in Fig.~\ref{fig:cavity}). With the output field operators introduced in Sec.~\ref{sec:io} at hand, one can readily calculate these observables as
\begin{align}
 R &= \frac{\ew{a_\text{out}}} {a_\text{in}} \label{eqn:reflectivity}\,, \\
 T &= \frac{\ew{b_\text{out}}} {a_\text{in}} \label{eqn:transmittance} \;.
\end{align}
Note that in current experiments, the reflected (transmitted) intensity $|R|^2$ ($|T|^2$) is measured, since phase information is often not accessible. By making use of an interferometric setup, also phase information could be retrieved.

Another observable of interest which can easily be accessed with the formalism developed here is the photon correlation function
\begin{equation}
 g^{(2)}(\tau) = \frac{\ew{a_\text{out}^\dagger(0) a_\text{out}^\dagger(\tau) a_\text{out}(\tau) a_\text{out}(0)}} {\ew{a_\text{out}^\dagger a_\text{out}}^2} \; .\label{eq:g2}
\end{equation}
It can be used to determine the photon statistics (at $\tau = 0$) as a function of any parameter or, if the operators are evaluated at different times ($\tau \neq 0$), photon \mbox{(anti-)}bunching~\cite{Scully_Zubairy_Quantum_Optics}. This way, quantum properties of the scattered light can be accessed. It should be noted that Eq.~(\ref{eq:g2}) characterizes temporal correlations between individual photons along the propagation direction of the scattered light, rather than spatial correlations in a transverse cross section through the propagating beam.

In this work we will focus on the reflectance $|R|^2$ calculated with Eq.~(\ref{eqn:reflectivity}) since it is of interest in current experiments.

\subsection{\label{sec:inc_nuc}Inclusion of the resonant nuclei}

\begin{table}
\begin{tabular}{c|c|c|c|c}
 $\quad\mu\quad$ & $\quad$Transition$\quad$ & $\quad\Delta E\quad$& $\quad$C-G$\quad$ & Polarization \\
 \hline
 1 & $\ket{g_1}\leftrightarrow\ket{e_1}$ & $-\delta_g/2-3/2\delta_e$ & $1$ & $\sigma^-$ \\
2 & $\ket{g_1}\leftrightarrow\ket{e_2}$ & $-\delta_g/2-1/2\delta_e$ & $\sqrt{2/3}$ & $\pi^0$ \\
3 & $\ket{g_1}\leftrightarrow\ket{e_3}$ & $-\delta_g/2+1/2\delta_e$ & $\sqrt{1/3}$ & $\sigma^+$ \\
4 & $\ket{g_2}\leftrightarrow\ket{e_2}$ & $\delta_g/2-1/2\delta_e$ & $\sqrt{1/3}$ & $\sigma^-$ \\
5 & $\ket{g_2}\leftrightarrow\ket{e_3}$ & $\delta_g/2+1/2\delta_e$ & $\sqrt{2/3}$ & $\pi^0$ \\
6 & $\ket{g_2}\leftrightarrow\ket{e_4}$ & $\delta_g/2+3/2\delta_e$ & $1$ & $\sigma^+$
\end{tabular}
\caption{Overview of the $M1$ allowed transitions in the $^{57}$Fe nucleus with transition index $\mu$. Shown are the involved states, the transition energy $\Delta E$ relative to the energy at vanishing magnetization $\omega_0$, the Clebsch-Gordan coefficient (CG) $c_\mu$ and the polarization type. Linear polarization is denoted by $\pi^0$, right (left) circular polarization as $\sigma^+$ ($\sigma^-$).}
\label{tab:transitions}
\end{table}

So far, we formulated the equations for an empty cavity. Next, we include the resonant nuclei. In general the nuclei have a multi-level structure, as discussed in Sec.~\ref{sec:nuclei}. But before we consider the general case with a magnetic hyperfine splitting, let us first consider the simplest case of a single two-level nucleus with ground state $\ket g$, excited state $\ket e$ and transition energy $\omega_0 = \omega_e - \omega_g$ and only one cavity mode $a$. This amounts to omitting the polarization dependence in this first step. In the Schr\"odinger picture the free time evolution of the nucleus and its coupling to the cavity mode in rotating wave approximation can be written as~\cite{Scully_Zubairy_Quantum_Optics}
\begin{align}
 H_N^{(S)} = \omega_g \ketbra{g} + \omega_e \ketbra{e} + g S_+ a + g^* a^\dagger S_- \;. \label{eqn:Hamiltonian_nuclei_simple}
\end{align}
Here $S_+ = \ket e\bra g$ and $S_- = \ket{g}\bra{e}$ denote the nuclear raising and lowering operators, respectively, and $g$ is the coupling constant between the mode $a$ and the nucleus. In order to transform the Hamiltonian for both the nuclei and the cavity modes into an time-independent interaction picture we alter the transformation from Eq.~(\ref{eqn:how_to_trafo}) to
\begin{equation}
 H_T = \omega \, a^\dagger a + \omega_g \, \ketbra{g} +  (\omega_g + \omega)  \ketbra{e} \; .
\end{equation}
This yields
\begin{align}
 H_N^{(I)} = - \Delta \ketbra{e} + g S_+ a + g^* a^\dagger S_- \;.
\end{align}
Here, we defined the detuning $\Delta = \omega - \omega_0$ as the energy difference between the external x-ray field and the bare transition energy of the nucleus.

Now we will continue with the general case including a possible magnetic hyperfine splitting caused by a field $\vec{B}_\text{hf}$. When a ferromagnetically ordered layer of $\alpha$-iron is placed in the cavity, already a relatively weak external field can align a strong internal magnetization of $\approx 33$T, resulting in a level splitting of several linewidths $\gamma$. The energy difference between two adjacent ground (excited) sub-states are denoted by $\delta_g$ ($\delta_e$) in the following. For $B\approx 33$T the values of $\delta_g$ and $\delta_e$ are $39.7\gamma$ and $22.4\gamma$, respectively \cite{Hannon_Trammell_Coherent_gammaray_optics}.

Using a similar transformation as above, the free evolution of $N$ nuclei and their coupling to the cavity modes $a_1$ and $a_2$ is given by the Hamiltonian 
\begin{align}
 H_N = &\sum_{n=1}^N H_0^{(n)} + H_{C_1}^{(n)}  + H_{C_2}^{(n)}
\end{align}
with the diagonal part
\begin{align}
 H_0^{(n)} =& \sum_{j=1}^2 \delta_g (j- \tfrac{3}{2}) \; \ketbra{g_j^{(n)}} \nonumber \\
     +& \sum_{j=1}^4 \left( \delta_e (j- \tfrac{5}{2}) - \Delta \right) \; \ketbra{e_j^{(n)}} \,.
 \label{eqn:Hamiltonian_nuclei}
\end{align}
The coupling between the the $n$th atom and the mode $a_j$ reads
\begin{align}
  H_{C_j}^{(n)} = \sum_{\mu=1}^6 & \left[ ( \vec{\hat d}_\mu^* \cdott  \vec{\hat a}_j )\, g_\mu^{(n)} S_{\mu+}^{(n)} a_j  \right. \nonumber \\
   +& \left. ( \vec{\hat a}_j^* \cdott \vec{\hat d}_\mu )\, {g_\mu^{(n)}}^* a_j^\dagger S_{\mu-}^{(n)} \, \right ] \; ,
\end{align}
where the sums run over the six possible transitions (see Tab.~\ref{tab:transitions}). The operator $S_{\mu +}^{(n)}$ [$S_{\mu -}^{(n)}$] acts only on atom $n$ and is the raising [lowering] operator on transition $\mu$. The normalized dipole moment $\vec{\hat d}_\mu$ of transition $\mu$ is defined with respect to the quantization axis of the nuclei, i.e.~the orientation of the magnetic hyperfine field $\vec{\hat B}$. The coupling constant 
\begin{align}
g_\mu^{(n)} = g \, c_\mu \, e^{i\, \vec{k}_C\cdott\vec{R}^{(n)}}
\end{align}
consists of the coupling constant $g$, the Clebsch-Gordan coefficient $c_\mu$ of the transition and a phase factor depending on the position $\vec{R}^{(n)}$ of the nucleus.

Another contribution which has to be included in the description of the nuclei is spontaneous emission. It can take place on each of the six transitions $\mu$, weighted with their respective Clebsch-Gordan coefficients $c_\mu^2$. Spontaneous emission is described with the Lindblad operator \cite{Scully_Zubairy_Quantum_Optics,Kiffner_Vacuum_Processes}
\begin{subequations}
 \label{eqn:spont_emission}
\begin{align}
 \mathcal{L}_\text{SE}[\rho] &= \sum_{n=1}^N \mathcal{L}_\text{SE}^{(n)}[\rho] \\
 \mathcal{L}_\text{SE}^{(n)}[\rho] &= -\frac{\gamma}{2} \sum_{\mu=1}^{6} c_\mu^2 \, \mathcal{L}[\rho, S_{\mu+}^{(n)}, S_{\mu-}^{(n)}] \,
\end{align}
\end{subequations}
where $\mathcal{L}[\rho, \cdot, \cdot]$ is defined in Eq.~(\ref{eqn:lindblad_helper}). Note that the expressions in Eqs.~(\ref{eqn:spont_emission}) characterize the total line width of single nuclei. Therefore, the rate of spontaneous emission $\gamma$ is taken as the natural line width of the $^{57}$Fe nucleus, even though part of this line width arises from internal conversion rather than from radiative decay.

\subsection{The full model}
The full master equation including the equations of motion of the nuclei as well as for the photonic modes is
\begin{align}
 \frac{d}{dt}\rho = -i [ H_M + H_N, \rho] + \mathcal{L}_M[\rho] + \mathcal{L}_\text{SE}[\rho]\;.
 \label{eqn:mastereqn_full}
\end{align}
With this equation it is in principle possible to perform calculations for arbitrary settings. However, the size of the system's Hilbert space a priori is infinite, because in general arbitrary occupation numbers of the photon modes are possible. Restricting the maximum number of photons per mode considered in the calculation to $n_{ph}$, the Hilbert space still scales as $6^N (n_{ph} + 1)^2$ with $N$ being the number of nuclei in the cavity, which is impractically large to be solved efficiently even for relatively small $n_{ph}$. Here we therefore use a different ansatz to overcome the obstacle of the fast growing Hilbert space, which in addition provides more insight in the underlying physics as even analytic predictions can be made. To this end we apply two physically motivated approximations. First, we make use of the fact that for typical parameters, the dissipative dynamics dominates the cavity evolution, such that the occupation number of the photon modes in the cavity remains small. Then, these photonic modes can be adiabatically eliminated to obtain effective equations of motion for the nuclei only, as explained in detail in Sec.~\ref{sec:adiabatic_elimination}. Second, in the case of a weak probe field, i.e. in linear response, the system of $N$ nuclei can be transformed into a new basis where only few excited states are coupled to the ground state. As shown in sections \ref{sec:appl1} and \ref{sec:appl2}, relatively simple analytic expressions can be found for the reflection coefficient in this case.

\section{Effective master equation}

\subsection{Adiabatic elimination of the cavity modes}
\label{sec:adiabatic_elimination}
The thin film cavities which are used in typical experiments have a low quality factor $Q$ \cite{Roehlsberger_Lamb_Shift}, which corresponds to a large decay constant $\kappa$ [see Eq.~(\ref{eqn:cavity_decay})] in our model. As $\kappa$ is much larger than the atom-field coupling strength $g$, the dynamics of the modes $a_1$ and $a_2$ is mainly governed by fast dissipation, which is known as bad cavity regime \cite{meystre_elements_quatum_optics}. This allows us to adiabatically eliminate the modes. For this, we approximate $\tfrac{d}{dt}a_j = 0$. Starting with the Heisenberg equation of motion for the operator $a_j$
\begin{equation}
 \frac{d}{dt} a_j = i [H_M + H_N, a_j] - \kappa a_j
\end{equation}
we arrive at
\begin{equation}
 a_j = \frac{\sqrt{2 \kR} \ain (\vec{\hat a}_j^*\cdott\vec{\hat a}_\text{in}) - i \sum_{n,\mu} (\vec{\hat a}_j^*\cdott\vec{\hat d}_\mu) {g_\mu^{(n)}}^* S_{\mu-}^{(n)}}{\kappa + i\dc} \;.\label{eq:aj}
\end{equation}
Before we continue with the effective equations for the nuclei let us consider the reflection coefficient as defined in Eq.~(\ref{eqn:reflectivity}). Inserting the expressions Eq.~(\ref{eq:aj}) for $a_j$ yields
\begin{align}
 R &= \frac{\ew{\aout}}{\ain} = \left(\frac{2\kR}{\kappa + i\dc} -1 \right) \vec{\hat a}_\text{out}^* \cdott \vec{\hat a}_\text{in} \nonumber \\
   &-\frac{i}{\ain}\frac{\sqrt{2\kR}}{\kappa +i\dc} \sum_{n,\mu} \left( \vec{\hat a}_\text{out}^* \cdott \mathbb{1}_\perp \cdott \vec{\hat d}_\mu \right) {g_\mu^{(n)}}^* \ew{S_{\mu-}^{(n)}} \; .
 \label{eqn:reflectivity2}
\end{align}
Here, we defined $\mathbb{1}_\perp = \vec{\hat a}_1 \vec{\hat a}_1^* + \vec{\hat a}_2 \vec{\hat a}_2^* = \mathbb{1} - \vec{\hat k} \vec{\hat k}^*$. Note that this expression contains outer products rather than inner (scalar) products.
We see that the reflection coefficient consists of two contributions. Consistent with the matrix formalism~\cite{Roehlsberger_Scattering} we can identify the first term in Eq.~(\ref{eqn:reflectivity2}) with the electronic scattering contribution, which is isotropic. A particularly interesting case arises if the cavity is operated exactly in resonance with the guided mode, i.e., $\dc =0$. If in addition $\kappa = 2\kR$ is fulfilled, then the reflection originating from the cavity vanishes completely. The latter condition is known as {\it critical coupling} condition~\cite{Dayan_Photon_Turnstile}. If the total cavity decay rate is not matched to the in- and out-coupling of light from the cavity, then the {\it over- or undercritically coupled} regime is realized, in which the reflected light is not completely canceled on resonance. Experimentally, the coupling regime can be controlled, e.g., via the thickness of the topmost layer of the waveguide.
The second term in Eq.~(\ref{eqn:reflectivity2}) describes the contribution to the reflection which is due to the nuclei. This contribution is not isotropic or polarization-preserving in general, and can contribute even if the polarizations of the incident beam and the detected radiation are orthogonal to each other.

We now continue with the adiabatic elimination of the cavity modes. Having established expressions for the field operators $a_j$ and $a_j^\dagger$, they can be inserted into the master equation (\ref{eqn:mastereqn_full}) to obtain the effective equations of motion for the nuclei. For the coherent dynamics, we obtain the Hamiltonian
\begin{align}
 H_\text{eff} =&\sum_{n=1}^N \left( H_0^{(n)} + H_\Omega^{(n)} \right) + \sum_{n,m=1}^N H_\text{LS}^{(n,m)}
\end{align}
with free evolution $H_0^{(n)}$ as defined in Eq.~(\ref{eqn:Hamiltonian_nuclei}) and the new terms
\begin{align}
 H_\Omega^{(n)} &= \Omega \sum_\mu \bigl( \vec{\hat d}_\mu^* \cdott \mathbb{1}_\perp \cdott \vec{\hat a}_\text{in} \bigr) g_\mu^{(n)} S_{\mu+}^{(n)} \nonumber \\
& + \Omega^* \sum_\mu \bigl( \vec{\hat a}_\text{in}^* \cdott \mathbb{1}_\perp \cdott \vec{\hat d}_\mu \bigr) {g_\mu^{(n)}}^* S_{\mu-}^{(n)} \,,\label{eqn:HOmega_full} \\
 H_\text{LS}^{(n,m)} &= \dls \sum_{\mu,\nu} \bigl( \vec{\hat d}_\mu^* \cdott \mathbb{1}_\perp \cdott \vec{\hat d}_\nu \bigr) g_\mu^{(n)} {g_\nu^{(m)}}^* S_{\mu+}^{(n)} S_{\nu-}^{(m)} \label{eqn:HLS_full}
\end{align}
arising from the adiabatic elimination procedure with parameters
\begin{align}
 \Omega &= \frac{\sqrt{2 \kR} \ain}{\kappa + i\dc} \,,\\
 \dls  &= -\frac{\dc}{\kappa^2+\dc^2} \;.
\end{align}
The Hamiltonian $H_\Omega^{(n)}$ describes an effective coupling between ground and excited states for each atom $n$. As expected, the transition dipole moments are not coupled to the polarization of the external beam by a direct product, but the direction vectors are mediated via the tensor $\mathbb{1}_\perp$ which reflects the intermediate light propagation in the two eliminated modes. To analyze the effect of $H_\text{LS}^{(n,m)}$ we first consider the special case $n=m$ and $\mu = \nu$, i.e., operators for the same transition in the same atom. It can be seen that in this case, the product $S_{\mu+}^{(n)} S_{\nu-}^{(m)}$ reduces to an operator of the form $\ketbra{e}$ for atom $n=m$. Therefore, this term in the Hamiltonian is an energy shift, which can be interpreted as an additional AC-Stark or Lamb shift emerging from the coupling of the atom to the two modes in the cavity. The terms with $n\neq m$ involving the same transition in different atoms are known as dipole-dipole interactions \cite{Ficek_Swain,Kiffner_Vacuum_Processes} and lead to a collective Lamb shift~\cite{Scully_Collective_Lamb_Shift_Theory}. In the cases $\mu\neq \nu$, a coherent coupling between two different transitions emerge~\cite{Agarwal_Quantum_statistical_spontaneous_emission,Ficek_Swain,Kiffner_Vacuum_Processes}.

Apart from these Hamiltonian contributions, the adiabatic elimination also gives rise to incoherent dynamics beyond spontaneous emission as characterized by Eqs.~(\ref{eqn:spont_emission}). The total Lindblad operator is found as
\begin{align}
\label{eq:lc}
 \mathcal{L}_\text{eff}[\rho] = &\mathcal{L}_\text{SE}[\rho] + \mathcal{L}_\text{cav}[\rho]
\end{align}
with the new term
\begin{align}
 \mathcal{L}_\text{cav}[\rho] = &-{\zs} \sum_{n,m=1}^N \sum_{\mu,\nu=1}^6 \bigl( \vec{\hat d}_\mu^*\cdott \mathbb{1}_\perp \cdott \vec{\hat d}_\nu \bigr) \, g_\mu^{(n)} {g_\nu^{(m)}}^* \nonumber \\
 &\qquad\qquad\qquad\qquad\times \mathcal{L}[\rho, S_{\mu+}^{(n)}, S_{\nu-}^{(m)}]\,, \label{eqn:LCav_full}
\end{align}
and
\begin{align}
 \zs = & \frac{\kappa}{\kappa^2+ \dc^2} \;.
\end{align}
The contributions with $n=m$ and $\mu=\nu$ in Eq.~(\ref{eq:lc}) have the same form as those characterizing spontaneous emission. As we will find in Sec.~\ref{sec:appl1}, they lead to superradiance, i.e., an acceleration of the incoherent decay \cite{Roehlsberger_Lamb_Shift}. The terms with $n=m$ and $\mu\neq \nu$ are so called cross decay terms \cite{Ficek_Swain} give rise to an incoherent coupling between different transitions. Interestingly, these terms can lead to coherences \cite{Kiffner_Vacuum_Processes}. This will be discussed in more detail in Sec.~\ref{sec:appl2}.

In both the coherent and the incoherent additions arising from the adiabatic elimination, the dipole moments are not coupled via the usual free space scalar product $\vec{\hat d}_\mu^*\cdott \vec{\hat d}_\nu$, but by the form $\vec{\hat d}_\mu^*\cdott \mathbb{1}_\perp \cdott \vec{\hat d}_\nu$. We emphasize that this generally permits non-vanishing couplings between \textit{orthogonal} states, which is fundamentally different from the situation in free space \cite{Patnaik_Preselected_Polarization,Agarwal_SGC_anisotropic_vacuum}. This fact can be exploited to engineer a variety of different quantum optical level schemes as will be shown in Sec.~\ref{sec:appl2}.

\subsection{Linear response}
\label{sec:linear_response}
Current experiments employing the $14.4$ keV resonance line in $^{57}$Fe in thin film cavities are mostly performed at modern synchrotron light sources. However, as the source bandwidth is orders of magnitude larger than the narrow resonance line width of $^{57}$Fe, each synchrotron pulse typically provides on average less than one resonant photon. Thus, the driving field $\ain$ can be considered weak, which together with the moderate nucleus-cavity coupling justifies a calculation of the reflectance in linear response. Of course, this ansatz has to be revisited if future experiments are performed at an seeded x-ray free electron laser or x-ray free electron laser oscillator~\cite{XFELO_Proposal,XFELO_Performance} with thousands of resonant photons per pulse, or if better cavities could be designed.

Let us assume that the nuclei are initially in the collective ground state
\begin{equation}
 \ket G = \underbrace{\ket{g_1^{(1)}}\ldots\ket{g_1^{(N_1)}}}_{N_1}\underbrace{\ket{g_2^{(N_1+1)}}\ldots\ket{g_2^{(N)}}}_{N_2}
\end{equation}
where $|g_1\rangle$ and $|g_2\rangle$ denote the two magnetic sublevels of the ground state, and $N_i$ is the number of nuclei in ground state $|g_i\rangle$ ($i\in \{1,2\}$). Note that $N_1 + N_2 = N$, and at room temperature and in thermal equilibrium also $N_1 = N_2$, since the Boltzmann factor $\exp{(-\delta_g/ k_B T)}$ is approximately one. Nevertheless, for now we consider the general case and keep $N_1$ and $N_2$ variable. Further, we assume that due to the weak probe beam only one atom can be excited at a time and omit higher excited states. In addition, we neglect other collective ground states as the nuclei will not be redistributed due to the application of a weak probe field. We define the singly excited states
\begin{equation}
 \ket{E_{\mu_e}^{(n)}} = S_{\mu+}^{(n)} \ket{G} = \ket{g_1^{(1)}}\ldots\ket{e_{\mu_e}^{(n)}}\ldots\ket{g_2^{(N)}}\,,
\end{equation}
in which the $n$th atom has been excited on transition $\mu$. Further we define the timed Dicke state \cite{Hannon_Trammell_Coherent_gammaray_optics,Scully_Collective_Lamb_Shift_Theory} 
\begin{equation}
 \ket{E_{\mu}^{(+)}} = \frac{1}{\sqrt{N_{\mu_g}}} \sum_n^{N_{\mu_g}} e^{i\, \vec{k}_C\cdot\vec{R}^{(n)}} \ket{E_{\mu_e}^{(n)}} \; ,
\end{equation}
which characterizes the coherent superposition of all possible excitations of the nuclei after absorption of a photon on transition $\mu$ where $\mu_g$ [$\mu_e$] denote the state index of the ground [excited] state of the transition. Note that atoms in ground state $\ket {g_n}$ can only be excited along the transition $\mu$ if their initial ground state match, i.e.~$\ket{g_n}=\ket{g_{\mu_g}}$, otherwise $ S_{\mu+}^{(n)} \ket{G} = 0$. With these definitions the equations (\ref{eqn:HOmega_full}), (\ref{eqn:HLS_full}), (\ref{eqn:LCav_full}) in the subspace of $\le 1$ excitations simplify to
\begin{align}
 H_{\Omega} &= \Omega g \sum_\mu \bigl( \vec{\hat d}_\mu^* \cdott \mathbb{1}_\perp \cdott \vec{\hat a}_\text{in} \bigr) c_\mu \sqrt{N_{\mu_g}} \ket{E_\mu^{(+)}}\bra{G} +  \textrm{H.c.} \,,\\
 H_\text{LS} &= \dls |g|^2 \sum_{\mu,\nu} \bigl( \vec{\hat d}_\mu^* \cdott \mathbb{1}_\perp \cdott \vec{\hat d}_\nu \bigr) \,,\nonumber \\
 &\qquad\qquad\qquad \times c_\mu c_\nu \sqrt{N_{\mu_g} N_{\nu_g}} \ket{E_\mu^{(+)}} \bra{E_\nu^{(+)}} \\
 \mathcal{L}_\text{cav}[\rho] &= -\zs |g|^2 \sum_{\mu,\nu}\bigl( \vec{\hat d}_\mu^* \cdott \mathbb{1}_\perp \cdott \vec{\hat d}_\nu \bigr) c_\mu c_\nu \sqrt{N_{\mu_g} N_{\nu_g}} \nonumber \\
 &\qquad\qquad\times \mathcal{L} \Big[\rho,  \ket{E_\mu^{(+)}}\bra{G},  \ket{G}\bra{E_\nu^{(+)}} \Big] \label{eqn:LA_linear} \; .
\end{align}
In this basis only one ground and a maximum of six (collective) excited state are present. This reduced basis allows for a considerable simplification of the analytical calculations since also the reflection coefficient can be written in the reduced basis as
\begin{align}
 R &= \left(\frac{2\kR}{\kappa + i\dc} -1 \right) \vec{\hat a}_\text{out}^* \cdott \vec{\hat a}_\text{in} \nonumber \\
   &-\frac{i}{\ain}\frac{\sqrt{2\kR}}{\kappa +i\dc}  {g}^* \sum_{\mu} \left( \vec{\hat a}_\text{out}^* \cdott \mathbb{1}_\perp \cdott \vec{\hat d}_\mu \right) c_\mu \sqrt{N_{\mu_g}} \nonumber \\
 &\qquad\times \bra{E_\mu^{(+)}}\rho\ket{G} \; .
 \label{eqn:reflectivity3}
\end{align}
This is a remarkable result, since the complicated system of $N$ interacting nuclei and 2 cavity modes is now reduced to an effective single-particle problem without loss of generality within the applied approximations well justified at current experimental conditions. 

At this point we can compare our results to the previously introduced matrix formalism~\cite{Roehlsberger_Scattering}. In the latter formalism, scattering amplitudes for two transitions coupling to linearly and four transitions coupling to circularly polarized light enter. Also within our framework, we naturally obtain these six transitions. This analogy is expected, since both the matrix formalism and the special case constructed in this section are linear in the probing field. Another analogy exists in the couplings between the different transitions. As they are mediated via the tensor $\mathbb{1}_\perp$, it is easy to see that for (anti-) parallel or orthogonal orientation of $\vec{B}_\text{hf}$ with respect to $\vec{\hat k}$, the excited states split into distinct subsets which are not mutually coupled. This corresponds to the situation in which the scattering matrix in the matrix formalism decomposes as it can be written as a direct product of two eigenpolarizations~\cite{Roehlsberger_Scattering,heeg_sgc}.

To calculate the reflection coefficient in linear response we employ the following method. We set $\bra{G} \rho \ket{G} = 1$, as population redistributions only occur in second order of the probe field. Next, we consider the coherences $\bra{E_\mu^{(+)}}\rho\ket{G}$ which are directly coupled to the ground state via $H_\Omega$. These off-diagonal density matrix elements are the only ones which are non-vanishing in first order in the probing x-ray field $\Omega$. Their steady state is obtained from the equations of motion by the condition $\bra{E_\mu^{(+)}}\tfrac{d}{dt} \rho\ket{G} = 0$. The corresponding set of linear equations can be solved easily. Finally, the obtained steady state is inserted into Eq.~(\ref{eqn:reflectivity3}) to obtain the desired reflectance in linear response.

\section{Application to particular experimental settings}
\label{sec:appl}
In this Section, we apply the general formalism to two particular experimental setups studied recently, in order to demonstrate its capabilities and consistency with previous formalisms. 

\subsection{Unmagnetized $^{57}$Fe layer}
\label{sec:appl1}
In a first step, we apply our formalism to the simplest case without hyperfine splitting, i.e.~$\vec{B}_\text{hf}=0$ and therefore $\delta_g = \delta_e = 0$. In this case the result will be independent of the choice of the quantization axis. For simplicity we set $\vec{\hat \pi}^0 \parallel \vec{\hat a}_\text{in}$ such that only the linear polarized transitions $\mu=2$ ($\ket{g_1}\leftrightarrow\ket{e_2}$) and $\mu=5$ ($\ket{g_2}\leftrightarrow\ket{e_3}$) are driven, see Tab.~\ref{tab:transitions}. We introduce the state 
\begin{align}
\ket + = \sqrt{\frac{N_1}{N}} \ket{E_2^{(+)}} + \sqrt{\frac{N_2}{N}} \ket{E_5^{(+)}}\,,
\end{align}
and obtain
\begin{subequations}
\begin{align}
 H_\Omega &= \sqrt{\tfrac{2}{3}N}\Omega g \ket{+}\bra{G} +  \textrm{H.c.} \label{eqn:HOmega_noB} \\
 H_\text{LS} &= \tfrac{2}{3} N \dls |g|^2 \ketbra{+} \label{eqn:HLS_noB} \\
 \mathcal{L}_\text{cav}[\rho] &= -\tfrac{2}{3} N \zs |g|^2  \, \mathcal{L}\Big[\rho,\ket{+}\bra{G}, \ket{G}\bra{+} \Big] \label{eqn:LA_noB} \;.
\end{align}
\end{subequations}
Thus, we have transformed our system to an effective two-level system which consists only of one ground state $\ket G$ and one excited state $\ket +$. In the same way, the sum in Eq.~(\ref{eqn:reflectivity3}) reduces to $(\vec{\hat a}_\text{out}^* \cdott \vec{\hat a}_\text{in})\sqrt{\tfrac{2}{3} N} \bra{+}\rho\ket{G}$. Consequently, only the coherence $\bra{+}\rho\ket{G}$ has to be calculated. The equation of motion is
\begin{align}
 \bra{+}&\dot\rho\ket{G} = -i \sqrt{\tfrac{2}{3} N}\Omega g \nonumber \\
 &+ i \left( \Delta + i\tfrac{\gamma}{2} + \tfrac{2}{3}N|g|^2 ( i\zs - \dls ) \right) \bra{+}\rho\ket{G} \;, \label{eqn:rhodot_plus_pi}
\end{align}
where we used the populations in linear response $\bra{G}\rho\ket{G} = 1$ and $\bra{+}\rho\ket{+} = 0$. Since $\bra{+}\rho\ket{G}$ is not coupled to any other density matrix elements in Eq.~(\ref{eqn:rhodot_plus_pi}), its steady state can be readily obtained from solving the single equation $\bra{+}\dot\rho\ket{G} = 0$ for the coherence $\bra{+}\rho\ket{G}$. The reflection coefficient given by Eq.~(\ref{eqn:reflectivity3}) evaluates to
\begin{align}
 R &= \left(\frac{2\kR}{\kappa + i\dc} -1 \right) \vec{\hat a}_\text{out}^* \cdott \vec{\hat a}_\text{in} \nonumber \\
   &-\frac{i}{\ain}\frac{\sqrt{2\kR}}{\kappa +i\dc} \frac{\bigl(\vec{\hat a}_\text{out}^*\cdott \vec{\hat a}_\text{in} \bigr) \tfrac{2}{3}|g|^2 N \Omega}{\Delta + i\tfrac{\gamma}2 + \tfrac{2}{3}|g|^2 N (i \zs - \dls)} \;.
  \label{eqn:reflectivity_no_B}
\end{align}
As a cross check we verified that this result is also obtained when choosing a different quantization axis, such that other transitions couple to the incident light. The result, however, in general does depend on the condition $N_1 = N_2$ of equal ground state population, as otherwise the different transitions have different probabilities according to the ratio of $N_1$ and $N_2$. The polarization dependence $\vec{\hat a}_\text{out}^* \cdott \vec{\hat a}_\text{in}$ is independent of the layer system, and solely determined by the incident and the detection polarization. This is the expected result, as no direction in space is distinguished in the layer system without magnetic quantization axis.

In order to interpret the spectrum of the reflectance, we first recall that the first addend obtained in Eq.~(\ref{eqn:reflectivity_no_B}) represents the electronic scattering contribution from the waveguide, while the second addend arises from the nuclei. Defining the parameters
\begin{subequations}
\begin{align}
 \Dls &= \frac{2}{3} \dls |g|^2 N\,,\\
 \gs &= \frac{4}{3} \zs |g|^2 N \;,
\end{align}
\end{subequations}
the nuclear part of the reflection coefficient can be rewritten as
\begin{align}
 R_\text{nuclei} \sim \frac{1}{\Delta - \Dls + \tfrac{i}{2} (\gamma + \gs)} \; .
\end{align}
This shape is a Lorentzian which describes the response of an effective two-level system with transition frequency shifted by $\Dls$ and spontaneous emission enhanced by $\gs$. Consistent with our theoretical modeling, the two levels correspond to the collective ground and the collective excited state of the nuclear ensemble. Note that even though $g$ is very small, the parameters $\Dls$ and $\gs$ will generally be of importance due to the large number of nuclei $N \gg 1$.

The adiabatic elimination of the cavity modes revealed couplings between the nuclei mediated by the cavity, such that collective effects emerge. The spontaneous emission enhancement $\gs$ is the well-known superradiance, and the energy shift $\Dls$ is a collective Lamb shift, both experimentally observed in~\cite{Roehlsberger_Lamb_Shift}. We see that both quantities contain contributions depending on cavity parameters ($\dls$, $\zs$). These can be related to the Purcell effect \cite{Purcell_effect}, which is the enhancement of spontaneous emission due to the cavity environment. The other contributions describes the cooperative behavior, as evidenced by the scaling with $N$.

\begin{figure}[t]
 \centering
 \includegraphics[scale=1]{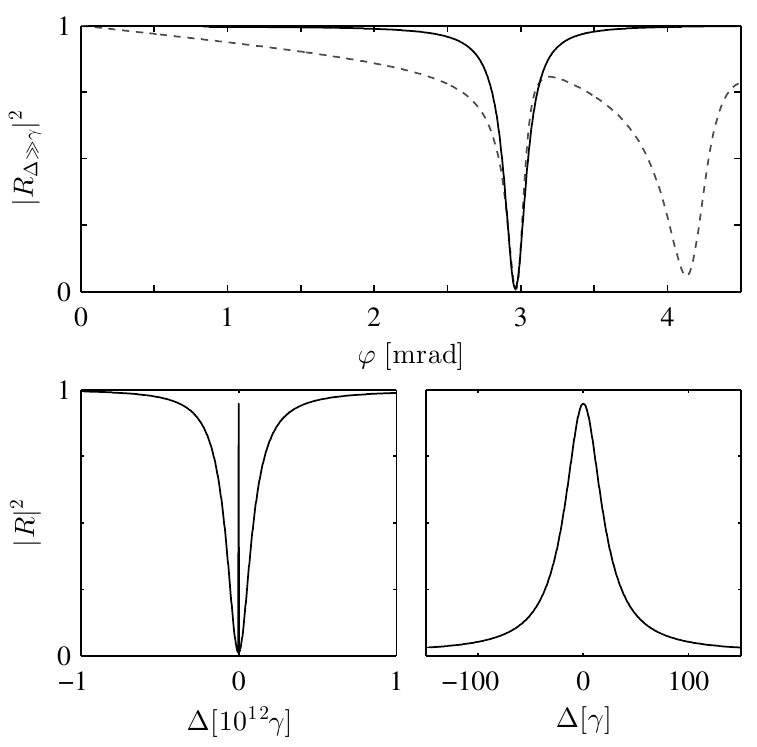}
 \caption{(Color online) Reflectance of the cavity containing an unmagnetized $^{57}$Fe layer. The top panel shows $|R|^2$ as a function of the grazing incidence angle $\varphi$. The nuclear part is strongly detuned such that only the electronic reflectivity curve is visible. Parameters as in the main text and $\varphi_0 = 2.96$ mrad. The dashed line corresponds to the reflectivity curve from Fig.~\ref{fig:rocking} calculated by CONUSS. The lower left panel shows the reflectance for fixed $\varphi = \varphi_0$. The narrow nuclear resonance is located in the center of the broad cavity resonance, where it appears as a sharp spike. The lower right panel shows a magnification of the lower left panel around the nuclear resonance. The nuclear spectrum is a Lorentzian which is significantly broadened due to superradiance.}
 \label{fig:R_no_B}
\end{figure}

At this point is instructive to discuss the actual values of the cavity parameters $\kappa$, $\kR$, $\dc$ and the coupling coefficient $g$. From the structure of Eq.~(\ref{eqn:reflectivity_no_B}) for the reflection coefficient, we note that the final result will be invariant under a rescaling $\xi$ of the parameters $\kappa$, $\kR$, $\dc$ and $N|g|^2$. Using numerical data
calculated by CONUSS~\cite{Sturhahn_CONUSS} for the cavity considered in Fig.~\ref{fig:rocking} as a reference, we find that, consistent with our expectations from Eq.~(\ref{eqn:cavity_detuning}), $\dc$ depends on the actual angle of incidence $\varphi$ in the vicinity of the first order guided mode fulfilling the relation $\dc = \delta_C \cdot \Delta\varphi$, while all other parameters remain constant. In particular, we find the values (in units of $\gamma$) $\kappa = 45 \xi$, $\kR = 25 \xi$, $\delta_C = -0.5\xi/\mu\text{rad}$ and $\sqrt{N}|g| = \sqrt{1400\xi}$. By comparison of $\dc = \delta_C \cdot \Delta\varphi$ with Eq.~(\ref{eqn:cavity_detuning}), the actual value for the scaling factor can be determined as $\xi \approx 18000$. Note that this also justifies the adiabatic elimination in Sec.~\ref{sec:adiabatic_elimination} since using the obtained parameters, we find $\kappa \gg \sqrt{N}|g|$.

The reflectance $|R|^2$ calculated from Eq.~(\ref{eqn:reflectivity_no_B}) is shown in Fig.~\ref{fig:R_no_B}. 
Note that the data shown in Fig.~\ref{fig:R_no_B} does not contain any free scaling parameter, as the reflectance calculated from Eq.~(\ref{eqn:reflectivity_no_B}) automatically yields the experimentally accessible values in the range between 0 and 1.
In the upper panel we qualitatively recover the shape of a typical electronic reflectivity curve across a single cavity resonance. To this end, we chose $\varphi_0 = 2.96$ mrad which is also the angle of the first guided mode in Fig.~\ref{fig:rocking}. In addition, we set the detuning $\Delta = 10^3 \gamma$ such that the nuclear part of the reflection is strongly suppressed and only the electronic part contributes. As a reference, we also show corresponding numerical results obtained with CONUSS.
It is clearly visible that in the vicinity of the first guided mode our theory matches the numerical data calculated with CONUSS very well. Since we included only one guided mode in our calculation, only one minimum in the reflectivity curve is obtained instead of multiple dips in the CONUSS data. Also, an overall envelope of the reflection, which in reality drops to smaller values for angles larger than the critical angle of total reflection, is not included in the theory, but visible in the CONUSS data. We emphasize that in our theory the width of the guided mode depends on the order of the scaling parameter $\xi$. But since $\xi$ was derived independently using Eq.~(\ref{eqn:cavity_detuning}), the proper width and the agreement with the numerical data serves also as a consistency check for our theory.

We now turn to the spectrum $|R(\Delta)|^2$ at the cavity resonance, i.e.~$\varphi=\varphi_0$. We find that a variation of the detuning $\Delta = \omega - \omega_0$ also affects the cavity detuning $\dc$, since it depends on both $\omega$ and $\omega_0$ explicitly [see Eq.~(\ref{eqn:cavity_detuning})]. Therefore we rewrite $\dc$ as a function of $\Delta$ and other constant parameters and show the results in the bottom panel of Fig.~\ref{fig:R_no_B}. In the bottom left panel we observe that the guided mode formed by the cavity affects the spectrum over a very large detuning range. Only in the center we observe the effect of the embedded nuclei, where the typical Lorentzian line shape of the nuclear resonance is found. A magnification of this nuclear response is shown in the bottom right panel. As expected from the theoretical predictions, in contrast to the resonance curve of a single $^{57}$Fe nucleus in free space, it is significantly broadened due to superradiance and the Purcell effect captured in $\gs$. We conclude from our analysis that if one is only interested in spectral ranges several $10\gamma$ around the nuclear resonance, it is safe to assume that $\Delta_C$ is independent of $\Delta$. The reason is that for any given angle $\varphi$, the cavity forms a nearly perfect flat background over the range of the nuclear response, as seen from Fig.~\ref{fig:R_no_B}.

\subsection{Magnetized $^{57}$Fe layer}
\label{sec:appl2}
Next, we include the magnetic hyperfine splitting in our analysis. In contrast to the calculation in the last section, then the six collective excited states $\ket{E_\mu^{(+)}}$ are no longer degenerate and thus multiple resonances in the spectrum of the reflectance are expected. Furthermore, since the magnetization distinguishes one direction in space, the rotational invariance observed in the results for the unmagnetized layer will break down.
If the transitions were independent from each other, the nuclear part of the reflection coefficient would be the sum of the respective Lorentz curves. However, this is not the case here as the transitions are mutually coupled via $H_\text{LS}$ and $\mathcal{L}_\text{cav}[\rho]$. These couplings depend on the orientation of $\vec{\hat B}_\text{hf}$. Moreover, the incidence and detection polarizations $\vec{\hat a}_\text{in}$ and $\vec{\hat a}_\text{out}$ influence the obtained spectra in a non-trivial way. Therefore, we expect significant deviations in the spectra from a naive sum of Lorentzians and a strong dependence on the relative orientation of the axes $\vec{\hat B}_\text{hf}$, $\vec{\hat a}_\text{in}$ and $\vec{\hat a}_\text{out}$. Both of these expectations were recently confirmed experimentally~\cite{heeg_sgc}.

\begin{figure}
 \includegraphics[scale=1]{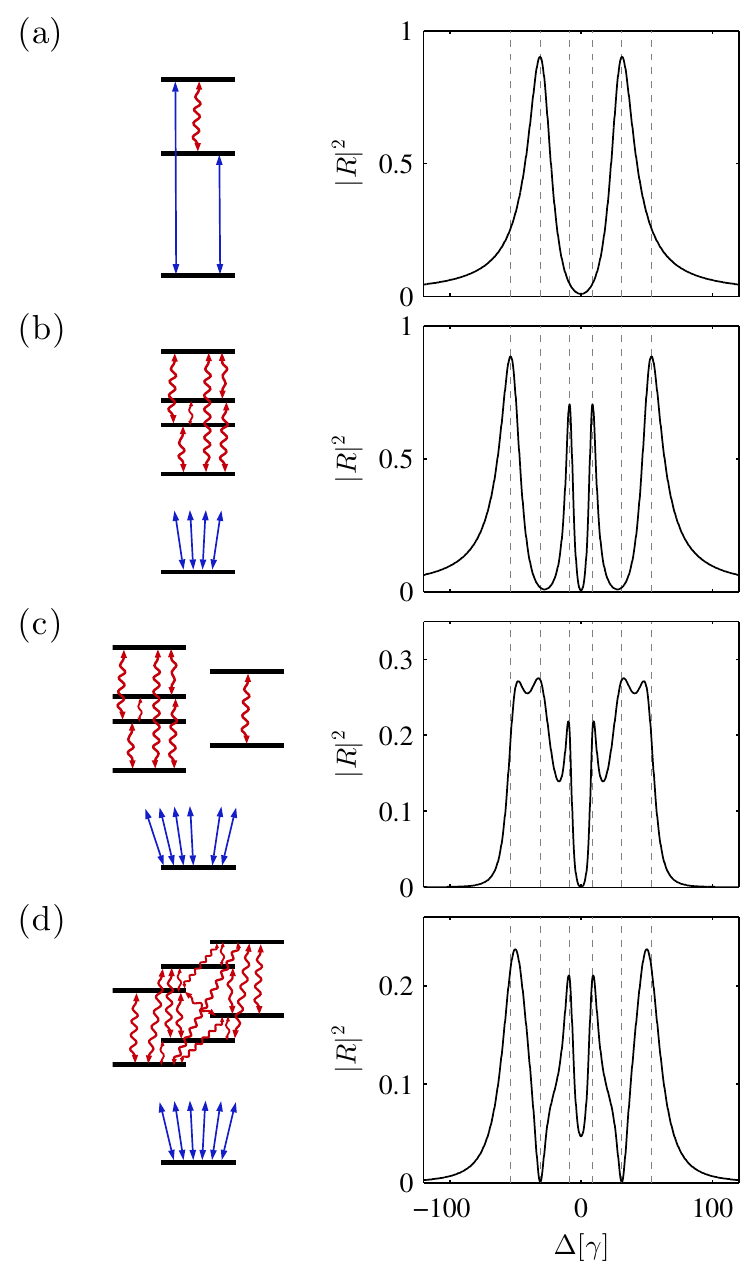}
 \caption{(Color online) 
 Engineering of nuclear level schemes. Depending on the choice of the input polarization and the nuclear magnetization axes, different level schemes are obtained. The four rows show the cases
(a)~$\vec{\hat a}_\text{in} \!\parallel\! \vec{\hat a}_\text{out} \!\parallel\! \vec{\hat B}_\text{hf}$,
(b)~$\vec{\hat a}_\text{in} \!\parallel\! \vec{\hat a}_\text{out} \!\perp\! \vec{\hat B}_\text{hf}$,
(c)~$\vec{\hat a}_\text{in} \!\parallel\! \vec{\hat a}_1\!-\!\vec{\hat a}_2$, $\vec{\hat a}_\text{out} \!\parallel\! \vec{\hat a}_1\!+\!\vec{\hat a}_2$, $\vec{\hat B}_\text{hf} \!\parallel\! \vec{\hat a}_2$,
and (d)~$\vec{\hat a}_\text{in} \!\parallel\! \vec{\hat a}_1$, $\vec{\hat a}_\text{out} \!\parallel\! \vec{\hat a}_2$, $\vec{\hat B}_\text{hf} \!\parallel\! \vec{\hat a}_2 \!+\! \vec{\hat k}$. 
The left column shows the obtained level scheme, and the right column the corresponding reflectance.
The excited states $\ket{E_\mu^{(+)}}$ are mutually coupled due to $H_\text{LS}$ and $\mathcal{L}_\text{cav}$ (red curly arrows) and coherently probed by $H_\Omega$ (blue). Spontaneous decay channels and Lamb shifts are not shown in the level diagram for clarity. The two geometries in (c,d) correspond to situations considered in \cite{heeg_sgc} where only the nuclear part of $R$ contributes. The vertical lines in the reflectance plots indicate the resonance frequencies of the six transitions. Other parameters are as in Fig.~\ref{fig:R_no_B}.}
 \label{fig:R_with_B}
\end{figure}

Effective level schemes for different choices of the polarization and magnetization alignment are shown in Fig.~\ref{fig:R_with_B}. The number of excited states and, equally important, their respective couplings induced by the cavity modes are modified considerably. This indicates that a vast range of different quantum optical level schemes can be engineered in a single sample, only by suitably choosing the different polarization and magnetization axes.
Accordingly, also the reflectances differ from each other as it can be seen in the right panel of Fig.~\ref{fig:R_with_B}.
The most prominent features are the peaks at the respective resonance energies of the different transitions, indicating the multi-level structure of the level schemes. But in addition, also repeatedly occurring minima are found in all spectra. As shown in \cite{heeg_sgc}, these minima are caused by the presence of so-called spontaneously generated coherences~\cite{Agarwal_Quantum_statistical_spontaneous_emission,Ficek_Swain,Kiffner_Vacuum_Processes}, which lead to vanishing spectral response due to destructive interference.

To understand the origin of the spectra in more detail, we consider the simplest case of $\vec{\hat B}_\text{hf} \parallel \vec{\hat a}_\text{in} \parallel \vec{\hat a}_\text{out}$ shown in Fig.~\ref{fig:R_with_B}(a). Here, only the linearly polarized transitions are driven. For simplicity we use $N_1 = N_2 = N/2$ in the following. Similar to the analysis in the last section we introduce states
\begin{subequations}
\begin{align}
 \ket + &= \frac{1}{\sqrt{2}} \left( \ket{E_5^{(+)}} + \ket{E_2^{(+)}} \right)\,, \\
 \ket - &= \frac{1}{\sqrt{2}} \left( \ket{E_5^{(+)}} - \ket{E_2^{(+)}} \right) \;.
\end{align}
\end{subequations}
The Hamiltonian written in this basis is found from our general theory as 
\begin{align}
 H =& -\Delta \big( \ketbra{+} + \ketbra{-} \big) \nonumber \\
 &+\tfrac{1}{2}(\delta_g+\delta_e) \big( \ket{+}\bra{-} + \ket{-}\bra{+} \big) \nonumber \\
 &+ \left (\sqrt{\tfrac{2}{3}N}\Omega g \ket{+}\bra{G} + \textrm{H.c.} \right )\nonumber \\
 &+ \tfrac{2}{3} N \dls |g|^2 \ketbra{+} \;.
\end{align}
This form reveals that only the fully symmetric state $\ket +$ of all allowed singly excited states is driven by the applied probe field. But in contrast to the case without magnetic field, the symmetric state $\ket +$ is coupled to a different state $\ket -$ in the presence of the magnetic field splitting, such that now a system of two linear equations needs to be solved. For the full treatment one has to consider the decay of the two involved excited states in addition. It turns out that the density matrix element $\bra{+}\rho\ket{G}$ decays exponentially due to spontaneous emission and enhanced by superradiance with rate $\tfrac{1}{2}\gamma + \tfrac{2}{3} N \zs |g|^2$, while $\bra{-}\rho\ket{G}$ decays only with $\gamma/2$. Since the superradiant decay is much faster than intrinsic spontaneous emission, $\ket -$ is metastable on the evolution timescale of $\ket +$. The origin of the suppression of the decay lies in the special form of the incoherent dynamics in Eq.~(\ref{eqn:LA_linear}). Due to the presence of the cross decay terms (the parts with $\mu \neq \nu$) not the bare excited states $\ket{E_{2}^{(+)}}$ and $\ket{E_{5}^{(+)}}$, but the \mbox{(anti-)}symmetrized states $\ket +$ and $\ket -$ are the radiative eigenstates with respect to the total decay. Hence, the cross decay terms naturally induce a coherence between the excited states which is known as spontaneously generated coherence.

\begin{figure}[t]
 \centering
 \includegraphics[scale=1]{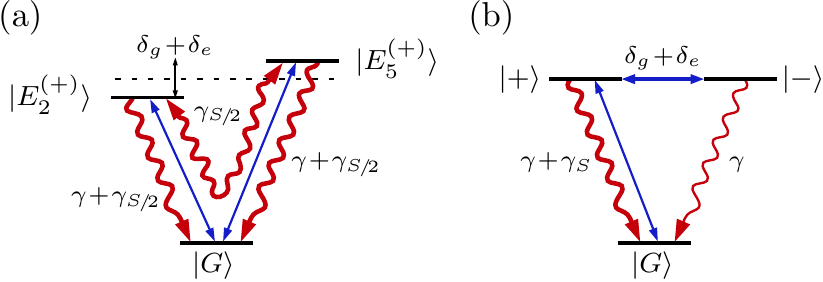}
 \caption{(Color online) The effective level system obtained if the linearly polarized transitions are driven by the probing field in the presence of a magnetic splitting. Collective Lamb shifts are not considered in the figure for clarity. (a) The collective ground state $\ket G$ is coherently coupled to the two possible excited states (solid blue arrows). Both states decay superradiantly (singly-headed red curly arrows) and are coupled via cross decay terms (double-headed curly arrow). (b) After a basis transition, only the symmetric state $|+\rangle$ is probed by the incident field. It is coupled to the antisymmetric state $|-\rangle$, which is metastable on the superradiantly accelerated decay time scale of $|+\rangle$ since it decays at the single-nucleus incoherent decay rate.}
 \label{fig:level_scheme_pi_sgc}
\end{figure}

The full level scheme for this particular orientation of polarizations and magnetization is shown in more detail in Fig.~\ref{fig:level_scheme_pi_sgc}. The complexity of the large ensemble of nuclei readily visible in the single-nucleus basis $\ket{E_\mu^{(+)}}$ is entirely hidden in the description with $\ket{+}$ and $\ket -$. In the latter basis, the nuclear ensemble can be identified with a typical $V$ or $\Lambda$ level scheme, as required for electromagnetically induced transparency (EIT) \cite{Fleischhauer_EIT_review,Roehlsberger_EIT,Schmidt_EIT_Comparison}. Therefore, it is clear that we rediscover the well known transparency dip from EIT also in the reflectance of our system.

The deep interference minima in other geometric realizations can be understood in a similar way.
The resulting analytic expressions for the nuclear part of the reflection coefficient are in perfect agreement with the prediction of the matrix formalism \cite{Roehlsberger_Scattering} and a previously used quantum optical description \cite{heeg_sgc}. Both these formalisms, however, have the disadvantage that analytic expressions for the reflectance could not readily be calculated for the cases where the quantization axis $\vec{\hat B}_\text{hf}$ and the beam propagation direction $\vec{\hat k}$ are either parallel or perpendicular. In other situations, such as the one shown in Fig.~\ref{fig:R_with_B}(d), a numerical study had to be performed. Furthermore, the different physical processes contributing to the obtained spectrum can not be distinguished. In contrast, our approach renders analytic calculations for general choices of the axes possible and agrees with the previous numerical results. In the general case all six excited states $\ket{E_\mu^{(+)}}$ need to be considered and thus a system of six equations has to be solved to obtain any observable in linear response. From the master equation, each physical process contributing to the final response can easily be identified and quantified. 

\section{Summary and Discussion}
We presented a quantum optical framework for thin film cavities containing layers of resonant nuclei, probed by hard x-rays in grazing incidence. This setting has recently been used in several experiments exploring the foundations of x-ray quantum optics.
Compared to previously existing frameworks, our approach allows for a full interpretation of all physical processes contributing to the observed signals, on the basis of a full understanding of the involved states and their mutual couplings from a microscopic point of view. In particular, we focused on the archetype M\"ossbauer isotope $^{57}$Fe which is presently also in the focus of interest in current experiments.
To overcome the difficulty of the large Hilbert space in the initial formulation of our theory, two well justified approximations were made. First, we adiabatically eliminated the cavity modes to obtain effective equations of motion for the nuclei. While there is no direct interaction among the nuclei initially, this procedure gives rise to mutual couplings in the equations. This way, an intuitive understanding of the relevant physical processes contributing to the coupling of the nuclei and thus to cooperative phenomena can be gained. In particular, we found that the cavity leads to an enhanced decay rate and energy shifts due to cooperativity and the Purcell effect. The second approximation was to consider the system only in first order of the driving field, which is sufficient to describe current experiments. Here, we found that cavity and the collective behavior of the nuclei can be described by one ground and up to only six collective excited states in the presence of a magnetic hyperfine splitting. It is important to note that these states are of excitonic nature, i.e., they are coherent superpositions of possible excitations to any of the nuclei in the whole nuclear ensemble. We note that it is the design of the cavity and the resulting geometrical arrangement of the nuclei in the cavity which enable the description in terms of collective states rather than individual atoms.

We then applied the formalism to particular settings of current experimental interest, and focused on the calculation of the x-ray reflectance. In the respective limits, we found excellent agreement with the previous models, as well as to numerical calculations using CONUSS. In the case of a plain cavity, we could recover the collective Lamb shift and, fundamentally linked, superradiant enhancement of spontaneous emission. These effects have been observed already in the considered cavity \cite{Roehlsberger_Lamb_Shift} as well as in the visible regime \cite{Keaveney_Coll_Lamb_Shift}. A more involved setting was studied by introducing a magnetic hyperfine splitting in the $^{57}$Fe nuclei. We found that a large set of level schemes can be engineered and controlled by suitable choices of the magnetization and polarization axes. This opens perspectives for the realization of advanced quantum optical level schemes with nuclei. In addition, we showed that interatomic interaction effects strongly modify the spectrum of the reflected signal, mainly due to the presence of vacuum induced coherences, as observed in \cite{heeg_sgc}. Within our framework, we could show that the resulting level scheme is analytically equivalent to the archetype electromagnetically induced transparency (EIT) setup \cite{Fleischhauer_EIT_review}. But in contrast to the usual EIT setting involving a coupling and a probe field, in our setting, only a probe field, but no externally applied coupling field is used. Rather, it is the effect of spontaneously generated coherences and the splitting of the upper levels, which effectively take this role \cite{Schmidt_EIT_Comparison}. Furthermore, cooperative effects are required to render one of the obtained excited states meta-stable on the time scale of the accelerated decay dynamics of the other states, in order to achieve EIT.

In this work we mainly applied our formalism to situations studied in synchrotron experiments, where the linear response approximation is valid. However, it is not limited to the linear regime and can in principle be applied to describe future experiments with much higher probe intensity studying non-linear effects, e.g., performed at seeded x-ray free electron lasers or x-ray free electron laser oscillators~\cite{XFELO_Proposal,ICFA}. Since in contrast to previous approaches we used a quantized field description, the accessible observables also cover more involved ones such as photon correlation functions. The model is suitable for any resonant nucleus with arbitrary hyperfine level structure. One could also easily extend it to multiple guided modes by including more cavity modes characterized by a certain wavenumber. Therefore, our formalism provides a promising platform for the further exploration of x-ray quantum optics with nuclei embedded in thin film waveguides.

\section*{Acknowledgements}
Fruitful discussions with R. R\"ohlsberger are gratefully acknowledged. K.P.H. acknowledges funding by the German National Academic Foundation.

\bibliographystyle{myprsty}
\bibliography{qomodel}
\end{document}